\documentclass{article}
\usepackage{amssymb}

\usepackage{amsmath}

\newtheorem{theorem}{Theorem}

\newtheorem{proposition}[theorem]{Proposition}

\input{tcilatex}

\begin{document}

\title{CLASSICAL \ AND \ QUANTUM\ \ SYSTEMS: ALTERNATIVE\ \ HAMILTONIAN\ \
DESCRIPTIONS}
\author{G. Marmo$^{1,3}\thanks{%
E-mail: marmo@na.infn.it}$, G. Scolarici$^{2}\thanks{%
E-mail: scolarici@le.infn.it}$, A. Simoni$^{1,3}\thanks{%
E-mail: simoni@na.infn.it}$ and F. Ventriglia$^{1,3,4}\thanks{%
E-mail: ventriglia@na.infn.it}$ \\
$^{1}$Dipartimento di Scienze Fisiche, Universita' di Napoli Federico II,
Complesso \\
Universitario di Monte S Angelo, Napoli, Italy.\\
$^{2}$Dipartimento di Fisica, Universita' di Lecce and INFN, Sezione di
Lecce, Italy\\
$^{3}$INFN, Sezione di Napoli, Italy\\
$^{4}$INFM, UdR di Napoli, Italy}
\maketitle

\begin{abstract}
In complete analogy with the classical situation (which is briefly reviewed)
it is possible to define bi-Hamiltonian descriptions for Quantum systems. We
also analyze compatible Hermitian structures in full analogy with compatible
Poisson structures.

\end{abstract}

\section{Introduction}

In the past thirty years a large number of nonlinear evolution equations
were discovered to be integrable systems \cite{gar}. It is a fact that
almost in all cases integrable systems also exhibit more than one
Hamiltonian descriptions, i.e. they admit alternative Hamiltonian
descriptions (they are often called bi-Hamiltonian systems) \cite{Carinena}.
In connection with quantum mechanics, there have been proposals for studying
complete integrability in the quantum setting.\cite{4},\cite{11}

If we take the view point of Dirac \cite{1}

\textsl{``Classical mechanics must be a limiting case of quantum mechanics.
We should thus expect to find that important concepts in classical mechanics
correspond to important concepts in quantum mechanics and, from an
understanding of the general nature of the analogy between classical and
quantum mechanics, we may hope to get laws and theorems in quantum mechanics
appearing as simple generalizations of well known results in classical
mechanics''},

it seems quite natural to ask the question: which alternative structures in
quantum mechanics, in the appropriate limit, will provide us with
alternative structures available in classical mechanics?

In particular, is it possible to exhibit the analog of alternative
Hamiltonian descriptions in the quantum framework?

This problem has been investigated in a two-pages paper by Wigner in
connection with commutation relations for the one-dimensional Harmonic
Oscillator.\cite{Wigner} See also Refs. \cite{Bloch}, \cite{Zaccaria}, \cite%
{8}, \cite{9}.

As we are interested in the structures rather than on specific applications,
it is better to consider the simplest setting in order to avoid
technicalities.

To clearly identify directions we should take in the quantum setting, it is
appropriate to briefly review the search for alternative Hamiltonian
descriptions in the classical setting, leaving aside the problem of
existence of compatible alternative Poisson brackets which would give rise
to complete integrability of the considered systems.

The paper is organized in the following way. In Section 2 we deal with
alternative Hamiltonian descriptions for classical systems, while in Section
3 the particular case of Newtonian equations of motion is addressed and in
Section 4 a meaningful example is discussed in detail. The analog picture in
Quantum case is exposed in Section 5 using the Weyl approach for the
Classical-Quantum transition. The Schroedinger picture is the framework to
study alternative descriptions of the equations of motion for Quantum
Systems in Section 6, in the finite dimensional case. The algebraic results
obtained there in the search for invariant Hermitian structures are extended
to infinite dimensions in the last part of the paper. In particular, in
Section 7, some theorems of Nagy are recalled to provide an invariant
Hermitian structure and in Section 8, starting with two Hermitian
structures, the group of bi-unitary transformations has been characterized
and a simple example is used to show how the theory works. Finally, some
concluding remarks are drawn in Section 9.

\section{Alternative Hamiltonian descriptions for classical systems}

Almost in all cases completely integrable classical systems are
bi-Hamiltonian systems. A dynamical system on a manifold $M$ is said to be
bi-Hamiltonian if there exists two Poisson Brackets denoted by $\left\{
.,.\right\} _{1,2}$ and corresponding Hamiltonian functions $H_{1,2}$ such
that

\begin{equation}
\frac{df}{dt}=\left\{ H_{1}{\Large ,}f\right\} _{1}=\left\{ H_{{\Large 2}}%
{\Large ,}f\right\} _{{\Large 2}}\text{ \ \ },\ \ \forall f\in \mathcal{F}%
(M).
\end{equation}
With any Poisson Bracket we may associate a Poisson tensor defined by 
\begin{equation}
\{\xi _{j},\xi _{k}\}=\Lambda _{jk}\ \ ,\ \ \Lambda =\Lambda _{jk}\frac{%
\partial }{\partial \xi _{j}}\wedge \frac{\partial }{\partial \xi _{k}}\ \ 
{\Large .}
\end{equation}

To search for alternative Hamiltonian descriptions for a given dynamical
system associated with a vector field $\Gamma $ on a manifold $M$, with
associated equations of motion 
\begin{equation}
\frac{df}{dt}=L_{\Gamma }f{\Large \ \ },
\end{equation}
we have to solve the following equation for the Poisson tensor $\Lambda $: 
\begin{equation}
L_{\Gamma }\Lambda =0\ \ .  \label{Poisson tensor}
\end{equation}
The vector field $\Gamma $ will be completely integrable if we can find two
Poisson tensors $\Lambda _{1}$ and $\Lambda _{2}$ , out of the possible
alternative solutions of \ equation (\ref{Poisson tensor}), such that any
linear combination $\lambda _{1}\Lambda _{1}+\lambda _{2}\Lambda _{2}$
satisfies the Jacobi identity. In this case the Poisson structures are said
to be compatible.\cite{Antonini}\ In particular, constant Poisson tensors $%
\Lambda _{1}$ and $\Lambda _{2}$ are compatible.

Summarizing, given a vector field $\Gamma $ 
\begin{equation}
\Gamma =\Gamma _{j}\frac{\partial }{\partial \xi _{j}}
\end{equation}
we search for pairs $(\Lambda ,H)$ which allow to decompose $\Gamma $ in the
following product 
\begin{equation}
\Gamma _{j}=\Lambda _{jk}\frac{\partial H}{\partial \xi _{k}}\ ,
\end{equation}
along with the additional condition (Jacobi identity): 
\begin{equation}
\Lambda _{jk}\partial _{k}\Lambda _{lm}+\Lambda _{lk}\partial _{{\Large k}%
}\Lambda _{mj}+\Lambda _{mk}\partial _{k}\Lambda _{jl}=0\ {\Large .}
\end{equation}

When the starting equations of motion are second order, further
considerations arise.

\section{Alternative Hamiltonian descriptions for equations of Newtonian type%
}

We recall that, according to Dyson,\cite{Dyson} ,\cite{Ibort} Feynman
addressed a similar problem, with the additional condition of
localizability; i.e. written in terms of positions and momenta $(x_{j},\
p_{j})${\Large \ } the localizability condition reads 
\begin{equation}
\{x_{j}{\Large ,}\text{ }x_{k}\}=0\ .
\end{equation}%
Thus, the search of Hamiltonian descriptions for a second order differential
equation reads 
\begin{equation}
\overset{\cdot }{x}_{j}=\{H,x_{j}\},
\end{equation}%
\ 
\begin{equation}
\overset{{\Large \cdot \cdot }}{x}_{j}=\{H,\{H,x_{j}\}\}=f_{j}(x,\overset{%
\cdot }{x})
\end{equation}%
Now we have to solve for the pair $(\{.,.\},H)$: it is clear that the
problem is highly non-trivial. However, if we require the localizability 
\begin{equation}
\{x_{j},x_{k}\}=0
\end{equation}%
and make the additional requirement (Galileian boost invariance) 
\begin{equation}
\frac{\partial }{\partial \overset{\cdot }{x_{j}}}\{.,\ .\}=0\ \ ,
\end{equation}%
we gain an incredible simplification.

Indeed starting with 
\begin{equation}
\overset{{\Large \cdot }}{x}_{j}=\{H,x_{j}\}
\end{equation}
and by taking the derivative with respect to $\overset{\cdot }{x}_{k}$, we
find 
\begin{equation}
\delta _{jk}=\frac{\partial ^{2}H}{\partial \overset{\cdot }{x_{k}}\partial 
\overset{\cdot }{x}_{m}}\{\overset{\cdot }{x}_{m},x_{k}\}\ .
\end{equation}
We have obtained that the bracket is not degenerate and the Hessian of $H$
is also not degenerate. We may now use a Legendre-type transformation to go
from the Hamiltonian description in terms of $H$ to the Lagrangian
description in terms of $\pounds $ and then the corresponding problem in
terms of Lagrangian functions is linearized and we have to solve for $%
\pounds $ the following equation

\begin{equation}
\frac{\partial ^{2}\pounds }{\partial \overset{{\Large \cdot }}{x}%
_{j}\partial \overset{\cdot }{x}_{m}}f_{m}(x,\overset{\cdot }{x})+\frac{%
\partial ^{2}\pounds }{\partial \overset{{\Large \cdot }}{x}_{j}\partial
x_{m}}\overset{{\Large \cdot }}{x}_{m}-\frac{\partial \pounds }{\partial
x_{j}}=0.
\end{equation}

Formulated in these terms the problem goes back to Helmholtz.\cite{Douglas}

\section{A Paradigmatic Example}

We shall consider a simple example that will be useful also to discuss the
corresponding quantum situation.\cite{act}

On \ $M=\mathbb{R}^{2n}$ , we consider 
\begin{equation}
\Gamma =\sum_{k=1}^{n}\lambda _{k}(p_{k}\frac{\partial }{\partial x_{k}}%
-x_{k}\frac{\partial }{\partial p_{k}})\ ,\ \lambda _{k}\in \mathbb{R}.
\end{equation}%
This $\Gamma $ represents the dynamical vector field of the anisotropic
Harmonic Oscillator with frequencies $\lambda _{k}$. As $\frac{\partial }{%
\partial p_{k}}\wedge \frac{\partial }{\partial x_{k}}$\ is invariant under
the flow associated with $\Gamma $ it follows that for any constant of the
motion $F(x,p)$ the following two-form is invariant, that is. 
\begin{equation}
L_{\Gamma }\omega _{F}=0
\end{equation}%
with {\Large \ } 
\begin{equation}
\omega _{F}=\sum_{k=1}^{n}\mu _{k}d(\frac{\partial F}{\partial p_{k}})\wedge
d(\frac{\partial F}{\partial x_{k}}),\ \ \ \mu _{k}\in \mathbb{R}\mathbf{.}
\end{equation}

For the one dimensional Harmonic Oscillator the function 
\begin{equation}
F(q,p)=(p^{2}+q^{2})(1+f(p^{2}+q^{2}))^{2}
\end{equation}
provides the most general invariant two-form parameterized by $%
f(p^{2}+q^{2}) $. For instance, 
\begin{equation}
F(q,p)=\exp \frac{\lambda }{2}(p^{2}+q^{2})
\end{equation}
gives 
\begin{equation}
\omega _{{\Large F}}=dP\wedge dQ,\ P=\lambda pF,\ Q=\lambda qF,
\end{equation}
with 
\begin{equation}
\{P,Q\}=\lambda ^{2}F^{{\Large 2}}(q,p)[1+\lambda (p^{2}+q^{2})]\{p,q\}
\end{equation}
exhibiting the Poisson bracket for the new variables expressed in terms of
the old ones and showing that the transformation is not canonical. Now we
have to stress that the equations of motion are linear in the new variables 
\begin{equation}
\frac{d}{dt}P=-Q\ \ ,\ \ \frac{d}{dt}Q=P\ \ ,
\end{equation}
in addition to the linearity in the old variables.

We have obtained that the equations are linear in two different coordinate
systems with a connecting coordinate transformation which is not linear. We
notice that in each coordinate system, say $(p,q)$ and $(P,Q),$ the
following tensor fields are preserved by the dynamical evolution 
\begin{equation*}
\ \ \omega =dp\wedge dq\ ,
\end{equation*}
\begin{equation}
g=dp\otimes dp+dq\otimes dq\text{ \ },\ \Delta =p\frac{\partial }{\partial p}%
+q\frac{\partial }{\partial q},\text{\ \ }
\end{equation}
and 
\begin{equation*}
\ \ \omega ^{\prime }=dP\wedge dQ\ ,
\end{equation*}
\ 
\begin{equation}
g^{\prime }=dP\otimes dP+dQ\otimes dQ\text{ \ },\ \text{\ }\Delta ^{\prime
}=P\frac{\partial }{\partial P}+Q\frac{\partial }{\partial Q}.
\end{equation}
In each set of coordinates we have alternative realizations of both the
linear inhomogeneous symplectic group, preserving the corresponding
symplectic structure, and of the linear rotation group. Their intersection
yields alternative realizations of the unitary group. All these linear
realization are not linearly related.

\emph{How to formulate an analog picture for the quantum case?}

At the classical level, the dynamical vector field $\Gamma $ is a derivation
for the associative algebra $\mathcal{F}(M)$ and a derivation for the binary
product associated\textsf{\ }with the Poisson Bracket.

Would it be possible to define alternative ''Lie brackets'' and consider a
similar approach also in the quantum setting?

Unfortunately this naive approach does not work, when the algebra is
associative maximally non-commutative, the Lie brackets compatible with the
associative product is necessarily proportional to the commutator, i.e. $%
\lambda (AB-BA)$. \cite{1},\cite{Gra-Ma}

Therefore to change the Lie bracket one has to change also the associative
product \cite{4}, \cite{5}, \cite{6}.

To have an idea on how to search for alternative descriptions for quantum
systems it is convenient to consider Weyl approach to quantization because
in this approach the symplectic structure plays a well identified role.

\section{Quantum systems in the Weyl Approach}

Given a symplectic vector space $(E,\omega )$, a Weyl system \cite{16}, \cite%
{Baez}, \cite{11} is defined to be a strongly continuous map from $E$ to
unitary transformations on some Hilbert space $\mathbb{H}$ : 
\begin{equation}
W:E\rightarrow U(\mathbb{H})
\end{equation}
with 
\begin{equation}
W(e_{1})W(e_{2})W^{\dagger }(e_{1})W^{\dagger }(e_{2})=e^{^{\frac{{\Large i}%
}{{\Large \hbar }}{\Large \omega (e}_{1}{\Large ,e}_{2}{\Large )}}}\mathbb{I}%
,
\end{equation}
with $\mathbb{I}$ the identity operator.

Thus a Weyl system defines a projective unitary representation of the
Abelian vector group $E$ whose cocycle is determined by the symplectic
structure.

The existence of Weyl systems for finite dimensional symplectic vector space
is exhibited easily and it amounts to the celebrated von Neumann's theorem
on the uniqueness of the canonical commutation relations.\cite{2},\cite%
{Mackey}

Consider a Lagrangian subspace $L$ and an associated isomorphism 
\begin{equation}
E\rightleftharpoons L\oplus L^{\ast }=T^{\ast }L\ \ .
\end{equation}%
On $L$\ we consider square integrable functions with respect to a Lebesgue
measure on $L$, a measure{\Large \ }invariant under translations. The
splitting of $E$ allows to define $e${\Large \ }$=(\alpha ,x)$ and set 
\begin{equation}
W((0,x)\Psi )(y)=\Psi (x+y),
\end{equation}%
\ \ 
\begin{equation}
W((\alpha ,0)\Psi )(y)=e^{{\Large i\alpha (y)}}\Psi (y),
\end{equation}%
\begin{equation}
x,y\in L\ ,\alpha \in L^{\ast },\Psi \in \emph{L}^{2}(L,d^{n}y);
\end{equation}%
it is obvious that $W(e)$ are unitary operators and moreover they satisfy
condition (27) with $\omega $ being the canonical one on $T^{\ast }L$.

The strong continuity allows to use Stone's theorem to get infinitesimal
generators $R(e)$ such that 
\begin{equation}
W(e)=e^{{\Large i}R(e)}{\Large \ \ \ }\forall e\in E\ 
\end{equation}
and $R(\lambda e)=\lambda R(e)$ for any $\lambda \in \mathbb{R}$.

When we select a complex structure on $E$ 
\begin{equation}
J:E\rightarrow E\ \ ,\ \ J^{2}=-1\ \ ,
\end{equation}
we may define ''creation'' and ''annihilation'' operators by setting 
\begin{equation}
a(e)=\frac{1}{\sqrt{2}}(R(e)+iR(Je)),
\end{equation}
\ 
\begin{equation}
a^{\dagger }(e)=\frac{1}{\sqrt{2}}(R(e)-iR(Je)).
\end{equation}
By using this complex structure on $E${\Large \ }we may construct an inner
product on $E$ as 
\begin{equation}
\left\langle e_{1},e_{2}\right\rangle =\omega (Je_{1},e_{2})-i\omega
(e_{1},e_{2}),
\end{equation}
therefore creation and annihilation operators are associated with a K\"{a}%
hler structure on $E.$\cite{Weil} The introduction of ''creation'' and
''annihilation'' operators is particularly convenient to relate alternative
descriptions on the Hilbert space (Fock space) with alternative descriptions
on the space of observables.

The Weyl map allows to associate automorphisms on the space of operators
with elements $S$ of the symplectic linear group acting on $(E,\omega )$ ,
by setting 
\begin{equation}
\nu _{S}(W(e))=W(Se)=U_{S}^{\dagger }W(e)U_{S}
\end{equation}
At the level of the infinitesimal generators of the unitary group, we have 
\begin{equation}
U_{S}^{\dagger }R(e)U_{S}=R(Se)
\end{equation}

\textbf{Remark:} As the relation defining $U_{S}$ is quadratic, one is
really dealing with the metaplectic group rather than the symplectic one. %
\cite{folland} However we shall not insist on this difference.

The Weyl map can be extended to functions on $T^{\ast }L\rightleftharpoons E$%
, indeed we first define the symplectic Fourier transform \cite{folland} of $%
f${\Large \ }$\in $ $\mathcal{F}(E)$ 
\begin{equation}
f(q,p)=\frac{1}{2\pi \hbar }\int \widetilde{f}(\alpha ,x)e^{\frac{i}{\hbar }%
(\alpha q-xp)}d\alpha dx
\end{equation}
and then associate with it the operator $\widehat{A}_{f}$ defined by 
\begin{equation}
\widehat{A}_{f}=\frac{1}{2\pi \hbar }\int \widetilde{f}(\alpha ,x)e^{\frac{i%
}{\hbar }(\alpha \widehat{Q}-x\widehat{P})}d\alpha dx.
\end{equation}
\emph{Vice versa}, with any operator $A$ acting on $\mathbb{H}${\Large \ }we
associate a function $f_{A}$ on the symplectic space $E${\Large \ }by setting%
{\Large \ } 
\begin{equation}
f_{A}(e)=TrAW(e)\ ;
\end{equation}
this map is called the Wigner map. When $A$ represents a pure state, i.e. a
rank-one projection operator, the corresponding function is the Wigner
function. A new product of functions may be introduced on $\mathcal{F}(E)$
by setting 
\begin{equation}
\left( f_{A}\ast f_{B}\right) (e)=TrABW(e).
\end{equation}
We thus find that alternative symplectic structures on $E$ give rise to
alternative associative products on $\mathcal{F}(E),${\Large \ }all of them
being not commutative.

The dynamics on $\mathcal{F}(E)${\Large \ }can be written in terms of this
non-commutative product as 
\begin{equation}
i\hbar \frac{df_{A}}{dt}=f_{H}\ast f_{A}-f_{A}\ast f_{H}\ \ \ .
\end{equation}
In this approach it is very simple to formulate the ''suitable limit'' to go
from the quantum descriptions to the classical description by noticing that
the limit 
\begin{equation}
\lim\limits_{\hbar \rightarrow 0}-\frac{i}{\hbar }(f_{A}\ast f_{B}-f_{B}\ast
f_{A})=\left\{ f_{A},f_{B}\right\}
\end{equation}
(when it exists) defines a Poisson bracket on $\mathcal{F}(E)$.

A different expression for this product involving the Poisson tensor $%
\Lambda $ is given by 
\begin{equation}
\left( f\ast g\right) (x,y)=f(x,y)e^{i\frac{\hbar }{2}(\frac{\overleftarrow{%
\partial }}{\partial x}\frac{\overrightarrow{\partial }}{\partial y}-\frac{%
\overleftarrow{\partial }}{\partial y}\frac{\overrightarrow{\partial }}{%
\partial x})}g(x,y)
\end{equation}
where as usual $\frac{\overleftarrow{\partial }}{\partial \xi }$ and $\frac{%
\overrightarrow{\partial }}{\partial \xi }$ act to the left and to the right
respectively \cite{17}, \cite{14}, \cite{15}, \cite{Manko}.

Now it is clear that, by using for instance the alternative Poisson brackets
we derived for the one-dimensional Harmonic Oscillator in Section 4, we may
write 
\begin{equation}
\left( f\ast g\right) (q,p)=f(q,p)e^{i\frac{\hbar }{2}(\frac{\overleftarrow{%
\partial }}{\partial q}\frac{\overrightarrow{\partial }}{\partial p}-\frac{%
\overleftarrow{\partial }}{\partial p}\frac{\overrightarrow{\partial }}{%
\partial q})}g(q,p)
\end{equation}
or 
\begin{equation}
\left( f\ast g\right) (Q,P)=f(Q,P)e^{i\frac{\hbar }{2}(\frac{\overleftarrow{%
\partial }}{\partial Q}\frac{\overrightarrow{\partial }}{\partial P}-\frac{%
\overleftarrow{\partial }}{\partial P}\frac{\overrightarrow{{\Large \partial 
}}}{\partial Q})}g(Q,P)\ \ \ .
\end{equation}
In this way we get two alternative associative products on $\mathcal{F}(E)$%
{\Large \ }both admitting $\Gamma $, the dynamical vector field of the
Harmonic Oscillator, as a derivation.

In the same sense for the Schroedinger picture, on the Hilbert space of
square integrable functions on the line, we may use either the Lebesgue
measure $dq$ invariant under translations generated by $\partial $/$\partial
q$, or 
\begin{equation}
dQ=\lambda d(q\exp \frac{\lambda }{2}(p^{2}+q^{2}))
\end{equation}
invariant under translations generated by $\partial $/$\partial Q$ .

Summarizing, by using the Weyl approach, we have been able to show that to
search for alternative Hamiltonian descriptions for quantum systems we may
look for alternative inner products on the space of states or alternative
associative products on the space of observables. In the coming sections we
shall investigate the existence of alternative descriptions in the
Schroedinger picture.

Preliminary results for alternative descriptions in the Heisenberg picture
are available in Ref. \cite{4}.

\section{Equations of motion for Quantum Systems and alternative descriptions%
}

Equations of motion in the carrier space of states are defined by the
Schroedinger equation (we set $\hslash =1$):

\begin{equation}
i\frac{d}{dt}\psi =H\psi .
\end{equation}

Here we shall first restrict ourselves to a finite $n$-dimensional complex
vector space $\mathbb{H}$. The dynamics is determined by the linear operator 
$H$. To search for alternative descriptions, we look for all scalar products
on $\mathbb{H}$ invariant under the dynamical evolution.

If we define $\Gamma :\mathbb{H}\rightarrow T\mathbb{H}$ to be the map $\psi
\rightarrow (\psi ,-iH\psi )$, we have to solve for $L_{\Gamma }h=0$, $h$
representing an unknown Hermitian structure on $\mathbb{H}$.

We notice that any $h$ on $\mathbb{H}$ defines an Euclidean metric $g$, a
symplectic form $\omega $ and a complex structure $J$ on the realification $%
\mathbb{H}^{R}$ of the complex space $\mathbb{H}$:

\begin{equation}
h(.,.)=:g(.,.)+ig(J.,.).
\end{equation}
The imaginary part of $h$ is a symplectic structure $\omega $ on the real
vector space $\mathbb{H}^{R}$:

\begin{equation}
\omega (.,.):=g(J.,.).
\end{equation}%
Thus any two of previous structures will determine the third one so defining
an \emph{admissible }triple $(g,J,\omega )$.

It is clear that $L_{\Gamma }h=0$ is equivalent to $L_{\Gamma }\omega =0$, $%
L_{\Gamma }g=0$, $L_{\Gamma }J=0$, so that we may solve for $L_{\Gamma }h=0$
by starting from $L_{\Gamma }\omega =0$.

To solve for this last equation we introduce the bi-vector field $\Lambda $
associated with Poisson Brackets defined by $\omega $ in the standard way. %
\cite{MSSV},\cite{EMS} The vector field $\Gamma $ will be factorized in the
form

\begin{equation}
\Gamma =\Lambda ^{lk}\frac{\partial f_{H}}{\partial \xi ^{k}}\frac{\partial 
}{\partial \xi ^{l}}.
\end{equation}
The matrix $\Lambda ^{lk}$ satisfies the following conditions 
\begin{equation}
\Lambda ^{lk}=-\Lambda ^{kl}\ \ \ ;\ \ \Lambda ^{lk}\partial _{k}\Lambda
^{rs}+\Lambda ^{rk}\partial _{k}\Lambda ^{sl}+\Lambda ^{sk}\partial
_{k}\Lambda ^{lr}=0\ \ \ .
\end{equation}

As $\Gamma $ is linear and $\Lambda ^{lk}$ is constant, $f_{H}$ must be
quadratic: $f_{H}=\frac{1}{2}\xi ^{k}H_{km}\xi ^{m}$ and therefore if we
write $\Gamma =A_{l}^{k}\xi ^{k}\frac{\partial }{\partial \xi ^{l}}$ we have
the necessary and sufficient condition for $\Gamma $ to be Hamiltonian in
the form

\begin{equation}
A=\Lambda H  \label{fact}
\end{equation}
and we get the following:

\begin{proposition}
\textbf{\ }All alternative Hamiltonian descriptions for $\Gamma $ are
provided by all possible factorization of $A$ into the product of a
skew-symmetric matrix $\Lambda $ times a symmetric matrix $H$. Moreover it
is easy to show that the following equivalences hold: 
\begin{equation*}
A\Lambda +\Lambda A^{tr}=0\Leftrightarrow L_{\Gamma }\Gamma =0,
\end{equation*}
\begin{equation*}
HA+A^{tr}H=0\Leftrightarrow L_{\Gamma }H=0,
\end{equation*}
\begin{equation}
\omega A+A^{tr}\omega =0\Leftrightarrow L_{\Gamma }\omega =0,
\end{equation}
where $\omega $ stays for the matrix representing the symplectic structure.
\end{proposition}

In Ref. \cite{Giordano} it is shown that a necessary condition for the
existence of such a factorization (eq.(\ref{fact})) for $A$ is that $%
TrA^{2k+1}=0$ \ $\forall k\in \mathbb{N}$.

By using the $(1-1)$ tensor field $T=A_{k}^{l}d\xi ^{k}\otimes \frac{%
\partial }{\partial \xi ^{l}}$ we can obtain pairwise commuting Hamiltonian
vector fields $T^{2k}(\Gamma )$ with $[T^{2k}(\Gamma ),T^{2r}(\Gamma )]=0$.

Assuming that we have found a factorization for $A$, say $A_{m}^{l}=\Lambda
^{ls}H_{sm}$, we may investigate the existence of an invariant Hermitian
structure $h$ on $\mathbb{H}$. In the case $\det A\neq 0,$ if $H_{sm}$ is
positive definite, we may use it as a metric tensor $g$ to define a scalar
(Euclidean) product on $\mathbb{H}^{R}$. Then we can write the polar
decomposition of the operator $A$: $A=J|A|$ where, as usual, $|A|$ is
defined as $\sqrt{A^{\dagger }A}$. Since $KerA=\varnothing $, $J$ is
uniquely defined and is $g$-orthogonal: $J^{\dagger }J=JJ^{\dagger }=1$.

$J$ has the following properties:

i) $J$ commutes with $A$ and $|A|$: this follows from the fact that $J=A%
\frac{1}{\sqrt{A^{\dagger }A}}=A\frac{1}{\sqrt{-A^{2}}}$ is a function of $A$
only;

ii) $J^{2}=-1$: this follows because $A=J|A|=A=|A|J$ while $%
-A=-J|A|=A^{\dagger }=J^{\dagger }|A|$, then multiplication by $J$ yields $%
|A|=-J^{2}|A|$ and from $KerA=\varnothing $ the stated result is obtained.

To deal with the degenerate case, $\det A=0$, additional work is needed and
can be found in Ref. \cite{Morandi}.

Having obtained an invariant complex structure $J$ it is now possible to
define an invariant Hermitian structure by using the invariant positive
definite symmetric matrix $H_{sm}$ and the complex structure $J$. All in all
we have proven the following:

\begin{proposition}
Any vector field $\Gamma $ which admits an Hamiltonian factorization into $%
\Lambda H$, preserves an Hermitian structure whenever the Hamiltonian
function $f_{H}$ is positive definite.
\end{proposition}

As a consequence, on finite dimensional complex vector spaces, quantum
evolutions are provided by Hamiltonian vector fields associated with
quadratic Hamiltonian functions, which are positive definite. Because each
Hamiltonian function gives rise to an Euclidean product, it is clear that $%
\Gamma $ is at the same time the generator of both a symplectic and an
orthogonal transformation, therefore the generator of a unitary
transformation.

Besides, the way $J$ has been constructed out of $A$ may be used to show %
\cite{Morandi} that the ($1-1)$ tensor field associated with the matrix $J$
satisfies the property $J(\Gamma )=-\Delta $, where $\Delta $ is the
Liouville vector field $\Delta =\xi ^{k}\frac{\partial }{\partial \xi ^{k}}$%
. By using the dilation $\Delta $ it is possible to write the quadratic
Hamiltonian function in the coordinate free form 
\begin{equation}
\frac{1}{2}h(\Delta ,\Delta )=\frac{1}{2}g(\Delta ,\Delta )=:\text{g }.
\end{equation}

At this point, in complete analogy with compatible Poisson structures\cite%
{Magri},\cite{Gel},\cite{Imm}, we may introduce and analyze a notion of
''compatible Hermitian structures'' \ or more precisely compatible triples $%
(g_{a},J_{a},\omega _{a})$ ; $a=1,2.$

We consider two admissible triples $(g_{a},J_{a},\omega _{a})$ ; $a=1,2$ on $%
\mathbb{H}^{R}$and the corresponding Hermitian structures $%
h_{a}=g_{a}+i\omega _{a}$. We stress that $h_{a}$ \ is a Hermitian form on $%
\mathbb{H}_{a}$ \ which is the complexification of $\mathbb{H}^{R}$ \emph{%
via }$J_{a}$ so that in general $h_{1}$and $h_{2}$ are not Hermitian
structures on the $\emph{same}$ complex vector space.

Moreover we consider the associated quadratic functions 
\begin{equation}
\text{g}_{1}=\frac{1}{2}g_{1}(\Delta ,\Delta ),\text{g}_{2}=\frac{1}{2}%
g_{2}(\Delta ,\Delta )
\end{equation}%
to which correspond vector fields $\Gamma _{1}$ and $\Gamma _{2}$ via $%
\omega _{1}$ and $\omega _{2}$ respectively.

\textbf{Definition }\textit{Two Hermitian structures} $h_{1}$ \textit{and} $%
h_{2}$ \textit{are said to be compatible if:} $\ $%
\begin{equation}
L_{\Gamma _{1}}h_{2}=L_{\Gamma _{2}}h_{1}=0.
\end{equation}%
\textit{Equivalently:} 
\begin{equation}
L_{\Gamma _{1}}\omega _{2}=L_{\Gamma _{2}}\omega _{1}=0;\text{ \ \ }%
L_{\Gamma _{1}}g_{2}=L_{\Gamma _{2}}g_{1}=0.
\end{equation}

\bigskip

From

\begin{equation}
L_{\Gamma _{2}}(i_{\Gamma _{1}}\omega _{1})=i_{[\Gamma _{2},\Gamma
_{1}]}\omega _{1}=0
\end{equation}
we find immediately that $[\Gamma _{2},\Gamma _{1}]=0$. Moreover,
remembering that a given symplectic structure $\omega $ defines the Poisson
bracket $\{f,g\}=\omega (X_{g},X_{f})$, being $i_{X_{f}}\omega =df$, we
derive also

\begin{equation}
\{\text{g}_{1},\text{g}_{2}\}_{1}=0\text{ \ \ and \ \ }\{\text{g}_{1},\text{g%
}_{2}\}_{2}=0,
\end{equation}
where $\{.,.\}_{1,2}$ is associated with $\omega _{1,2}$.

Out of the two compatible Hermitian structures on the real vector space $%
\mathbb{H}^{R}$ we have the following $(1-1)$ tensor fields: $%
G=g_{1}^{-1}\circ g_{2}$, $T=\omega _{1}^{-1}\circ \omega _{2}$ and $J_{1}$, 
$J_{2}$. These four $(1-1)$ tensor fields generate an Abelian algebra and
are invariant under $\Gamma _{1}$ and $\Gamma _{2}$. It is also possible to
prove that $G=J_{1}\circ T\circ J_{2}^{-1}=-J_{1}\circ T\circ J_{2}.$

The following properties are easy to derive

\begin{eqnarray}
g_{1}(Gx,y) &=&g_{1}(x,Gy)=g_{2}(x,y)\ ,  \notag \\
g_{2}(Gx,y) &=&g_{2}(x,Gy)=g_{1}^{-1}(g_{2}(x,.),g_{2}(y,.))\ ;
\end{eqnarray}
\begin{equation}
g_{1}(Tx,y)=g_{1}(x,Ty)\ ;\ g_{2}(Tx,y)=g_{2}(x,Ty)\ ;
\end{equation}
\begin{eqnarray}
g_{1}(x,J_{2}y)+g_{1}(J_{2}x,y) &=&0\ ;\ g_{1}(J_{2}x,J_{2}y)=g_{1}(x,y)\ , 
\notag \\
g_{2}(x,J_{1}y)+g_{2}(J_{1}x,y) &=&0\ ;\ g_{2}(J_{1}x,J_{1}y)=g_{2}(x,y)\ .
\end{eqnarray}

Thus we have found\cite{Morandi} that:

\begin{proposition}
The\ $(1-1)$\ tensor fields $G$, $T$, $J_{1}$\ and $J_{2}$\ are a set of
mutually commuting linear operators. $G$\ and $T$\ are self-adjoint while $%
J_{1}$\ and $J_{2}$\ are skew-adjoint with respect to both metric tensors
.and moreover there are orthogonal transformations for both metric tensors
\end{proposition}

Now we can consider the implications on the $2n$-dimensional vector space $%
\mathbb{H}^{R}$ coming from the existence of two compatible Hermitian
structures.

The space $\mathbb{H}^{R}$ will split into a direct sum of eigenspaces $%
\mathbb{H}^{R}=\bigoplus_{k}\mathbb{H}_{\lambda _{k}}^{R}$ where $\lambda
_{k}$ are the distinct eigenvalues of $G$.

According to our previous statements, the sum will be an orthogonal sum with
respect to both metrics, and in each $\mathbb{H}_{\lambda _{k}}^{R}$, $%
G=\lambda _{k}\mathbb{I}_{k}$ with $\mathbb{I}_{k}$ the identity matrix in $%
\mathbb{H}_{\lambda _{k}}^{R}$. Out of the compatibility condition $T$ \
will introduce a further orthogonal decomposition of each $\mathbb{H}%
_{\lambda _{k}}^{R}$ of the form

\begin{equation}
\mathbb{H}_{\lambda _{k}}^{R}=\bigoplus_{r}W_{\lambda _{k},\mu _{k,r}}
\end{equation}
where $\mu _{k,r}$ are distinct eigenvalues of $T$ in $\mathbb{H}_{\lambda
_{k}}^{R}$.

The complex structures commute in turn with both $G$ and $T$, therefore they
will leave each one $W_{\lambda _{k},\mu _{k,r}}$ invariant. Now we can
reconstruct, using $g_{\alpha }$ and $J_{\alpha }$, two symplectic
structures. They will be block-diagonal in the decomposition of $\mathbb{H}%
^{R}$ and on each subspace $W_{\lambda _{k},\mu _{k,r}}$ they will be of the
form

\begin{equation}
g_{1}=\lambda _{k}g_{2},\text{ \ \ }\omega _{1}=\mu _{k,r}\omega _{2}.
\end{equation}

Therefore in the same subspaces $J_{1}=g_{1}^{-1}\omega _{1}=\frac{\mu _{k,r}%
}{\lambda _{k}}J_{2}$. From $J_{1}^{2}=J_{2}^{2}=-1$ we get $(\frac{\mu
_{k,r}}{\lambda _{k}})^{2}=1$, hence $\mu _{k,r}=\pm \lambda _{k}$ and $%
\lambda _{k}>0$. The index $r$ can then assume only two values,
corresponding to $\pm \lambda _{k}$.

All in all we have proved the following:

\begin{proposition}
If two Hermitian structures $h_{1}=g_{1}+i\omega _{1}$, $h_{2}=g_{2}+i\omega
_{2}$\ are compatible, then the vector space $H^{R}$\ will decompose into
the double orthogonal sum: 
\begin{equation}
\bigoplus_{k=1,...,r,\alpha =\pm 1}W_{\lambda _{k},\alpha \lambda _{k}}
\label{decomposizione}
\end{equation}
where the index $k=1,...,r\leq 2n$\ labels the eigenspaces of the $(1-1)$\
tensor $G=g_{1}^{-1}\circ g_{2}$\ corresponding to its distinct eigenvalues $%
\lambda _{k}>0$, while $T=\omega _{1}^{-1}\circ \omega _{2}$\ will be
diagonal with eigenvalues $\pm \lambda _{k}$\ on $W_{\lambda _{k},\pm
\lambda _{k}}$, on each of which 
\begin{equation}
g_{1}=\lambda _{k}g_{2},\text{ \ \ }\omega _{1}=\pm \lambda _{k}\omega _{2},%
\text{ \ \ }J_{1}=\pm J_{2}.
\end{equation}
As neither symplectic form is degenerate, the dimension of each one of $%
W_{\lambda _{k},\pm \lambda _{k}}$\ will be necessarily even.
\end{proposition}

At this point from two admissible triples $(g_{a},J_{a},\omega _{a})$ ; $%
a=1,2$ on $\mathbb{H}^{R}$ we can consider the corresponding Hermitian
structures $h_{a}=g_{a}+i\omega _{a}$. We stress that $h_{a}$ \ is a
Hermitian form on $\mathbb{H}_{a}$ \ which is the complexification of $%
\mathbb{H}^{R}$ \emph{via }$J_{a}$ so that in general $h_{1}$and $h_{2}$ are
not Hermitian structures on the $\emph{same}$ complex vector space. When the
triples are compatible, the decomposition of the space in eq. (\ref%
{decomposizione}) holds so that $\mathbb{H}^{R}$ can be decomposed into the
direct sum of the spaces$\ \mathbb{H}_{R}^{+}$ and $\mathbb{H}_{R}^{-}$ on
which $J_{1}=\pm J_{2},$ respectively. The comparison of $h_{1}$and $h_{2}$
requires a fixed complexification of $\mathbb{H}^{R}$, for instance $\mathbb{%
H}_{1}=\mathbb{H}_{1}^{+}\oplus \mathbb{H}_{1}^{-}$. Then using \
orthonormal basis \{$e_{k^{+}}\}$ and \{$e_{k^{-}}\}$ we can write 
\begin{equation*}
h_{1}(x,y)=\sum\limits_{k^{+}}x_{k^{+}}^{\ast
}y_{k^{+}}+\sum\limits_{k^{-}}x_{k^{-}}^{\ast }y_{k^{-}}\ \ \ ,
\end{equation*}
\begin{equation}
h_{2}(x,y)=\sum\limits_{k^{+}}\lambda _{k^{+}}x_{k^{+}}^{\ast
}y_{k^{+}}+\sum\limits_{k^{-}}\lambda _{k^{-}}x_{k^{-}}y_{k^{-}}^{\ast }\ \
\ .  \label{scalarprod}
\end{equation}
It is apparent that $h_{2}$ is not a Hermitian structure as it is neither
linear nor antilinear on the whole space $\mathbb{H}_{1}$.

Now it is possible to consider the case of a field $\Gamma $ which leaves
invariant both the compatible triples. As a \ result, the direct sum
decomposition of the space in eq. (\ref{decomposizione}) is invariant under
the action of $\ \Gamma .$ Moreover the field $\ \Gamma $ is generator of
both bi-orthogonal and bi-symplectic transformations on $\mathbb{H}^{R}$,
therefore generator of unitary transformations on $\mathbb{H}_{a}$ , $a=1,2$.

\section{Searching for invariant Hermitian structures}

In this section we would like to investigate the equation 
\begin{equation}
L_{\Gamma }h=0
\end{equation}
when the carrier space is some infinite dimensional Hilbert space.

As it is well known, in many physical instances, the dynamical vector field $%
\Gamma $ entering Schroedinger equation is associated with unbounded
operators. It follows that the search for solutions of $L_{\Gamma }h=0$ is
plagued with difficult domain problems. It is convenient therefore to search
for Hermitian structures solutions of the equation

\begin{equation}
\phi _{t}^{\ast }h=h,\text{ \ \ }\forall t\in \mathbb{R},  \label{flusso}
\end{equation}%
i.e. for Hermitian structures invariant under the one parameter group of
linear transformations describing the dynamical evolution.

We may consider, in more general terms the following problem: Given an
invertible transformation $T:\mathbb{H}\rightarrow \mathbb{H}$, under which
conditions there exist an invariant Hermitian structure $h$ such that

\begin{equation}
h(x,y)=h(Tx,Ty).
\end{equation}

As it is well known, in infinite dimensions the topology of the vector space
of states is an additional ingredient which has to be given explicitly. We
therefore assume that $\mathbb{H}$ is a Hilbert space with some fiducial
Hermitian structure $h_{0}$ , in general not invariant under the action of $%
T $. We do require $T$ to be continuous, along with its inverse, in the
topology defined by $h_{0}$ or by any other Hermitian structure,
topologically equivalent to $h_{0},$which allows us to consider bounded sets.

In the search for invariant Hermitian structures on $\mathbb{H,}$
topologically equivalent to $h_{0},$ we have this preliminary result:

\begin{proposition}
If $h_{1}$and $h_{0}$define the same topology on $\mathbb{H}$, there exists
a selfadjoint positive definite bounded operator $Q$\ such that 
\begin{equation}
h_{1}(x,y)=h_{0}(Qx,Qy).
\end{equation}
\end{proposition}

\textbf{Proof.} In order to $h_{1}$ and $h_{0}$ define the same topology on $%
\mathbb{H}$, it is necessary that there exist two real positive constants $%
A,B$ such that

\begin{equation*}
Ah_{1}(x,x)\leq h_{0}(x,x)\leq Bh_{1}(x,x),\text{ \ \ }\forall x\in \mathbb{H%
}\text{.}
\end{equation*}

The use of the Riesz theorem on bounded linear functionals immediately
implies that there exists a bounded positive and selfadjoint (with respect
to both Hermitian structures) operator defined implicitly by the equation

\begin{equation*}
h_{1}(x,y)=h_{0}(Gx,y),\text{ \ \ }\forall x,y\in \mathbb{H}.
\end{equation*}
The positiveness of $G$ implies $G=Q^{2}$ and the thesis follows at once.

Now we are ready to state few results which go back to B. Sz. Nagy.\cite%
{Nagy},\cite{Markov}

We first discuss when a flow $\Phi (t)$ is unitary with respect to some
Hermitian structure $h_{\Phi }$ which is a solution of eq. (\ref{flusso}).

In other words we shall establish conditions for eq. (\ref{flusso}) to have
solutions and as by-product we exhibit how to find some of them when some
appropriate conditions are satisfied.

Consider an automorphism $T$ of a Hilbert space $\mathbb{H}$ with a
Hermitian scalar product $h_{0}$, construct the orbits

\begin{equation}
\left\{ T^{k}\psi \right\} {\Large \ \ \ \ };\ \ \ k\in \{0,\pm 1,\pm
2,\dots \}
\end{equation}
and require that all of them, with respect to the norm induced by $h_{0}$,
are bounded sets for any $\psi $. The use of the principle of uniform
boundedness \cite{Simon} shows that this is equivalent to require that $T$
is uniformly bounded.

We recall that the automorphism $T$ on $\mathbb{H}$ is said to be uniformly
bounded if there exists an upper bound $c${\Large \ }$<\infty $ such that 
\begin{equation}
||T^{k}||<c\ \ ;\ \ \ k\in \{0,\pm 1,\pm 2,\dots \}\ \ .
\end{equation}

For such an operator the following theorem \cite{Nagy} holds:

\textbf{Theorem.(Bela de Sz. Nagy)} \emph{For a uniformly bounded operator }$%
T$\emph{\ \ there exists a bounded selfadjoint transformation }$Q$\emph{\
such that} 
\begin{equation*}
\frac{1}{c}I\leq Q\leq cI
\end{equation*}%
\emph{and }$QTQ^{-1}=U$\emph{\ is unitary with respect to the fiducial }$%
h_{0}$\emph{. This implies that }$T=Q^{-1}UQ$\emph{\ is unitary with respect
to} 
\begin{equation*}
h_{T}(\varphi ,\psi ):=h_{0}(Q\varphi ,Q\psi )\ .
\end{equation*}%
\textbf{Proof. \ }(Sketch) Define the invariant scalar product $%
h_{T}(\varphi ,\psi )$ as the limit, for $n$ going to infinity, of $\ $%
\begin{equation*}
h_{0}(T^{n}\varphi ,T^{n}\psi )=:h_{n}(\varphi ,\psi ).
\end{equation*}%
This is the limit of a bounded sequence of complex numbers which does not
exist in general, at least in the usual sense. Therefore use the generalized
concept of limit for bounded sequence, introduced by Banach and Mazur. \cite%
{Banach} This generalized limit\ (denoted as $Lim$ ) amounts to define the
invariant scalar product $h_{T}$ as the transformed scalar product $h_{n}$
''at infinity'', where $T$ is interpreted as the generator of a $\mathbb{Z}-$%
action on $\mathbb{H}$.

The same approach can be used \cite{Nagy} to deal with an $\mathbb{R}-$%
action instead of the $\mathbb{Z}-$action so that:

\textbf{Theorem.}\emph{When the one-parameter group of automorphisms }$\Phi
(t)$\emph{\ is uniformly bounded, that is } 
\begin{equation*}
\left\| \Phi (t)\right\| <c\ \ ,\ \ t\in (-\infty ,\infty )\ ,
\end{equation*}
\emph{there exists a bounded selfadjoint transformation }$Q$\emph{\ such
that }$Q\Phi (t)Q^{-1}=U(t)$\emph{\ is a one-parameter group of unitary
transformations or }$\Phi (t)$ \emph{is unitary with respect to. } 
\begin{equation*}
h_{\Phi }(.,.)=h_{0}(Q.,Q.)\ .
\end{equation*}

\bigskip

\textbf{Example.} As a simple example\cite{Nagi2} consider the group of
translation on the line realized on $L_{2}(\mathbb{R})$ with a measure which
is not translationally invariant, i.e. 
\begin{equation}
(T_{t}\Psi )(x):=\Psi (x+t)\ ,\ \Psi \in L_{2}(\mathbb{R},\rho (x)dx),
\end{equation}%
where $\rho (x)$ is any function $0<\alpha <\rho (x)<\beta <\infty $ and
denote by $h_{\rho }$ the corresponding scalar product. If the limit $%
\lim\limits_{x\rightarrow -\infty }\rho (x)$ exists , say $%
\lim\limits_{x\rightarrow -\infty }\rho (x)=a,$ then it is trivial to
compute the Banach limit because it agrees with a limit in the usual sense.
In fact by Lebesgue Theorem we have: 
\begin{equation*}
\lim\limits_{t\rightarrow \infty }\int\limits_{\mathbb{R}}\Psi ^{\ast
}(x+t)\Phi (x+t)\rho (x)dx=\lim\limits_{t\rightarrow \infty }\int\limits_{%
\mathbb{R}}\Psi ^{\ast }(x)\Phi (x)\rho (x-t)dx=
\end{equation*}%
\begin{equation}
=\int\limits_{\mathbb{R}}\lim\limits_{t\rightarrow \infty }\Psi ^{\ast
}(x)\Phi (x)\rho (x-t)dx=a\int\limits_{\mathbb{R}}\Psi ^{\ast }(x)\Phi
(x)dx=ah_{0}(\Psi ,\Phi ).
\end{equation}%
This shows that the Banach limit gives $h_{T}(\Psi ,\Phi )=ah_{0}(\Psi ,\Phi
),$ i.e.\ it is a multiple of the standard translation invariant scalar
product. Therefore 
\begin{equation}
h_{T}(\Psi ,\Phi )=h_{\rho }(Q^{2}\Psi ,\Phi )=h_{\rho }((\sqrt{\frac{a}{%
\rho }})^{2}\Psi ,\Phi )\ \ \ ,
\end{equation}%
that is $Q=\sqrt{\frac{a}{\rho }}$ and 
\begin{equation}
(U_{t}\Phi )(x)=(QT_{t}Q^{-1}\Phi )(x)=\sqrt{\frac{\rho (x+t)}{\rho (x)}}%
\Phi (x+t)
\end{equation}%
is unitary in $L_{2}(\mathbb{R},\rho (x)dx).$

Having discussed few results on the existence of invariant Hermitian
structure we may now look at the problem of compatible Hermitian structures.

\section{Bi-unitary transformations' group: the infinite dimensional case}

In quantum mechanics the Hilbert space $\mathbb{H}$ is given as a complex
vector space, because the complex structure enters directly the Schroedinger
equation of motion. It is therefore natural to require that the two
admissible triples $(g_{1},J_{1},\omega _{1})$ and $(g_{2},J_{2},\omega
_{2}) $ share the same complex structure: $J_{1}=J_{2}=J$. As we have shown
this entails that the two triples are compatible and the corresponding
structures $h_{1}$and $h_{2}$ are Hermitian on the same complex space $%
\mathbb{H}$ .

These Hermitian structures are related by the operator $G$ used before which
is selfadjoint with respect to both structures. The operator $G$ generates a
weakly closed commutative ring and a corresponding direct integral
decomposition of the Hilbert space: \ 
\begin{equation}
\mathbb{H}=\int_{\Delta }\mathbb{H}_{\lambda }d\sigma (\lambda ),
\end{equation}
where $\Delta $ is the spectrum of the positive bounded and selfadjoint
operator $G$ and $d\sigma $ is the corresponding measure.\cite{Naimark}

As $G$ acts as a multiplicative operator on each component space $\mathbb{H}%
_{\lambda },$ a straightforward generalization of the results of the finite
dimensional case eq.(\ref{scalarprod}) follows: in fact the forms of $h_{1}$%
and $h_{2}$ on $\mathbb{H}$ are:

\begin{equation*}
h_{1}(\varphi ,\psi )=\int\nolimits_{\Delta }<\varphi ,\psi >_{_{\lambda
}}d\sigma (\lambda )\ \ ,
\end{equation*}
\begin{equation}
h_{2}(\varphi ,\psi )=\int\nolimits_{\Delta }\lambda <\varphi ,\psi
>_{_{\lambda }}d\sigma (\lambda )
\end{equation}
where $<\varphi ,\psi >_{_{\lambda }}$is the inner product on the component $%
\mathbb{H}_{\lambda }$.

As a result, bi-unitary transformations are: 
\begin{equation}
U\varphi =\int_{\Delta }U(\lambda )\varphi _{\lambda }\ d\sigma (\lambda ).
\end{equation}
where $U(\lambda )$ is a unitary operator on the component $\mathbb{H}%
_{\lambda }$. \cite{Lecce}

In particular, when $G$ is cyclic, each $\mathbb{H}_{\lambda }$ is one
dimensional and $U(\lambda )$ becomes a multiplication by a phase factor: %
\cite{Naim3} 
\begin{equation}
U\varphi =\int_{\Delta }e^{i\theta (\lambda )}\varphi _{\lambda }\ d\sigma
(\lambda ).
\end{equation}
Therefore in this case the group of bi-unitary transformation is
parametrized by the $\sigma -$measurable real functions $\theta $ on $\Delta
.$ This shows that the bi-unitary group may be written as 
\begin{equation}
U_{\theta }=e^{i\theta (G)}\ \ .
\end{equation}

\textbf{Example}: Particle in a box, a double case.

Consider the operator $G=1+X^{2}$ , with $X$ position operator, on $%
L_{2}([-\alpha ,\alpha ],dx)$. It is Hermitian with spectrum $\Delta
=[1,1+\alpha ^{2}]$. From the spectral family of $X:$%
\begin{equation*}
P(\lambda )f=\chi _{{\Large [}-\alpha ,\lambda {\Large ]}}f
\end{equation*}%
where $\chi _{\lbrack -\alpha ,\lambda ]}$ is the characteristic function of
the interval $[-\alpha ,\lambda ]$, we get the spectral family $%
P_{G}(\lambda )$of $G$: 
\begin{equation*}
P_{G}(\lambda )=P(\sqrt{\lambda -1})-P(-\sqrt{\lambda -1})\text{ \ \ }.
\end{equation*}%
Now $G$ does not have cyclic vectors on the whole $L_{2}([-\alpha ,\alpha
],dx)$, because if $f$ is any vector, $xf(-x)$ is not zero and orthogonal to
all powers $G^{n}f$. This argument fails on $L_{2}([0,\alpha ],dx)$, where $%
\chi _{\lbrack 0,\alpha ]}$ is cyclic. Analogously on $L_{2}([-\alpha
,0],dx),$so we get the splitting in 2 $G$-cyclic spaces 
\begin{equation*}
L_{2}[-\alpha ,\alpha ]=L_{2}[-\alpha ,0]\oplus L_{2}[0,\alpha ]\text{ \ \ }.
\end{equation*}

From $P_{G}$ and those ciclic vectors we obtain the measure 
\begin{equation*}
d\sigma (\lambda )=d\sqrt{\lambda -1}
\end{equation*}%
for the decomposition of the Hilbert space 
\begin{equation*}
\mathbb{H}=\int\nolimits_{[1,1+\alpha ^{2}]}\mathbb{H}_{\lambda }\ d\sigma
(\lambda )\text{ \ },
\end{equation*}%
where $\mathbb{H}_{\lambda }$ is one-dimensional for the particle in the $%
[0,\alpha ]$ box while is two-dimensional for the $[-\alpha ,\alpha ]$ box.

\section{Concluding remarks}

In this paper we have shown that in analogy with the classical situation it
is possible to define alternative Hermitian descriptions for quantum
equations of motion. We have not undertaken the analysis to use compatible
alternative Hermitian structures to study quantum completely integrable
systems, this step will require that our operator algebras are realized as
algebras of differential operators acting on subspaces of square integrable
functions defined on the real spectrum of a maximal set of commuting
operators to be identified as position operators.

In the quantum-classical transition that we have mentioned in the
introduction we should analyze why the complex structure that plays such a
relevant role in quantum mechanics does not show up in the classical limit.

These issues will be taken up elsewhere in connection with the
quantum-classical transition.

\end{document}